\documentclass{elsart41}
\usepackage{graphics}
\usepackage{graphicx}
\usepackage{amssymb}
\begin{document}

\begin{frontmatter}



\title{Electronic structure and the Fermi surface of ThRhIn$_5$ 
in comparison with uranium and transuranium compounds}

%

\author[AA]{Takahiro Maehira\corauthref{Name1}},
\author[BB]{Takashi Hotta}

\address[AA]{Faculty of Science, University of The Ryukyus,
Nishihara, Okinawa 903-0213, JAPAN}  
\address[BB]{Advanced Science Research Center,
Japan Atomic Energy Research Institute, Tokai, Ibaraki 319-1195, JAPAN}

\corauth[Name1]{Tel.: +81 98 895 8508, fax: +81 98 895 8508. \\
{\it Email address}: maehira@sci.u-ryukyu.ac.jp.}

\begin{abstract}

By using a relativistic linear augmented-plane-wave method,
we clarify energy band structure and the Fermi surfaces of
recently synthesized thorium compound ThRhIn$_5$.
We find several cylindrical Fermi surface sheets, which are
similar to those of CeTIn$_5$ (T=Ir and Co), PuTGa$_5$ (T=Co and Rh),
and AmCoGa$_5$.
We discuss such similarity among the compounds including rare-earth
or actinide ions with different $f$ electron numbers.

\end{abstract}

\begin{keyword}
Relativistic linear APW method \sep Fermi surface 
\sep Thorium compounds

\PACS    71.27.+a; 71.18.+y; 71.15.-m
\end{keyword}
\end{frontmatter}


In order to clarify $5f$ electron properties,
it is useful to investigate thorium compounds both from
experimental and theoretical viewpoints,
since it is possible to extract pure $5f$-electron properties
by subtracting the $5f$ contributions of thorium compounds
from those of uranium and transuranium materials.
In fact, the electronic structure of thorium compounds
has been investigated by band theory as proper reference
for the study of $5f$ electronic states of actinide compounds.

In this paper, we calculate energy band structure and Fermi surface
for ThRhIn$_5$ \cite{Matsuda} by applying
a relativistic linear augmented plane wave (RLAPW) method.
It belongs to the material group with HoCoGa$_5$-type crystal
structure, frequently referred to as ``115''.
We compare the present result with those for Ce- \cite{Maehira1},
Pu- \cite{Maehira2,Opahle}, and Am-115 compounds \cite{Maehira3}.

\begin{figure}[t]
\begin{center}
\includegraphics[width=0.45\textwidth]{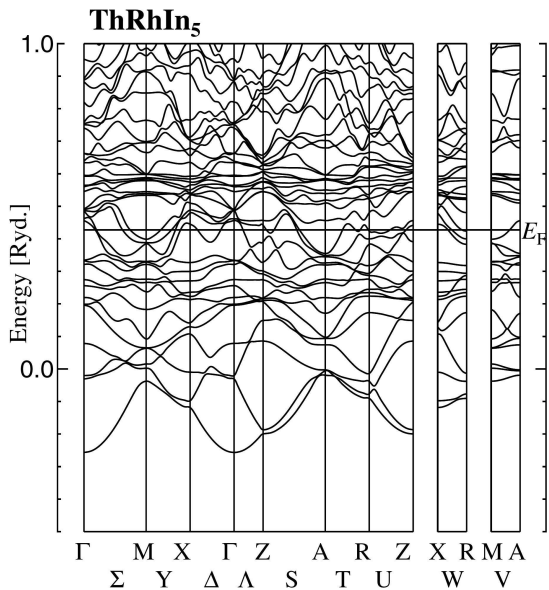}
\end{center}
\caption{Energy band structure of ThRhIn$_5$ calculated by the
RLAPW method. $E_{\rm F}$ indicate the position of the Fermi level.}
\label{fig1}
\end{figure}

Here we calculate the energy band structure
by using the RLAPW method with the exchange and
correlation potential in the local density approximation.
The muffin-tin approximation is adopted for the spatial shape
of one-electron potential and self-consistent calculations are
carried out for experimental lattice constants.
Note that all $5f$ electrons are assumed to be itinerant
in our calculations.
The energy band structure for ThRhIn$_5$ is shown in Fig.~\ref{fig1}.
The Fermi level, $E_{\rm F}$, is located at 0.428 Ryd. and shown
by a solid line in Fig.~\ref{fig1}.
Since 13th, 14th, 15th, and 16th bands are partially occupied,
these four bands construct the Fermi surface.
The number of the valence electrons in the APW sphere is partitioned
according to the sites and the angular momenta is as 
0.42(s), 6.13(p), 2.41(d) and 0.55(f) in the Th sphere.
There are 8.74 valence electrons outside the APW sphere
in the primitive cell.
Here we note a significant amount of $f$ component due to
large hybridization between $5f$ and $5p$ electron states.
The value of the $f$ component is a little larger than that of
fcc-Th, obtained by the RAPW calculation \cite{Higuchi},
in which the value of 0.41 was reported for the $f$ component.
The density of states (DOS) at $E_{\rm F}$ is evaluated as
56.6 states/Ryd.cell.
The contribution from the $f$ states to the DOS at $E_{\rm F}$
amounts to 14.8$\%$ of the total.
Above $E_{\rm F}$ near the M point, flat $5f$ bands split into two
groups, specified by $j$=5/2 for lower and 7/2 for upper bands,
respectively, where $j$ is the total angular momentum.
The magnitude of the splitting in $5f$ states
corresponds to the spin-orbit coupling, estimated as 0.048 Ryd.
As shown in Fig.~\ref{fig1}, the main part of the $f$ bands exist
just above $E_{\rm F}$.

\begin{figure}[t]
\begin{center}
\includegraphics[width=0.45\textwidth]{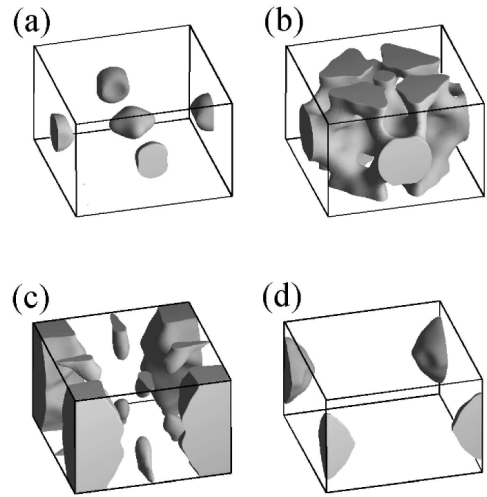}
\end{center}
\caption{Calculated Fermi surfaces of ThRhIn$_5$ for
(a) 13th band hole sheets, (b) 14th band hole sheets, 
(c) 15th band electron sheets, and (d) 16th band electron sheets.
Note that the center of the Brillouin zone is the $\Gamma$ point.}
\label{fig2}
\end{figure}

In Fig.~\ref{fig2}, we show the Fermi surfaces of ThRhIn$_5$.
The Fermi surfaces from the 13th band have one sheet centered at
the $\Gamma$ point, two equivalent sheets centered at the X points.
The 14th band constructs a large cylindrical hole sheet centered
at the $\Gamma$ point, which exhibits a complex network
consisting of big ``arms'' along the edges of the Brillouin zone,
as observed in Fig.~\ref{fig2}(b).
The 15th band has a cylindrical electron sheet centered at the M point.
The Fermi surface from the 16th band consists of one hole sheet
centered at the M point. 
The number of carriers contained in these Fermi-surface sheets are
0.051 holes/cell, 0.603 holes/cell, 0.581 electrons/cell,
and 0.073 electrons/cell in the 13th, 14th, 15th, and 16th bands,
respectively.
The total number of holes is equal to that of electrons,
since ${\rm ThRhIn_{5}}$ is a compensated metal.

We remark that the Fermi surfaces of ThRhIn$_{5}$ look
similar to those of CeTIn$_5$(T=Ir and Co) \cite{Maehira1},
PuTGa$_5$ (T=Co and Rh) \cite{Maehira2,Opahle},
and AmCoGa$_5$ \cite{Maehira3}.
If we assume the trivalent rare-earth or actinide ion
in 115 structure, we find zero, one, five, and six electrons
per ion for Th-, Ce-, Pu-, and Am-115 compounds, respectively.
In the $j$-$j$ coupling scheme, we accommodate $n$ electrons
in the $j$=5/2 sextet and thus, the electron-hole relation is
expected \cite{Hotta}.
In this sense, it seems to be natural to observe similar
Fermi-surface structure for the cases of $n$ and $6$$-$$n$
electrons.
Namely, it is enough to discuss the similarity between
the cases of $n$=0 and 1.
The Fermi surfaces of ThRhIn$_5$ constructed from
the four bands contain the $f$ components appreciably
due to the large hybridization between $f$ and $p$
electrons near the Fermi level.
In fact, in our band-calculation results,
we have found 0.55 $5f$ electrons in Th atom.
Then, the Fermi-surface structure of Th-115
becomes similar to that of Ce-115.

In summary, we have performed the band calculation for ThRhIn$_5$
by using the RLAPW method.
The similarity in the Fermi-surface structure among
Th-, Ce-, Pu-, and Am-115 compounds are understood by
the electron-hole relation in the $j$-$j$ coupling scheme
and the large hybridization between $f$- and
$5p$-electron states in the vicinity of $E_{\rm F}$.

We thank Y. Haga, K. Kubo, T. D. Matsuda, H. Onishi, Y. \=Onuki,
K. Ueda, H. Yamagami, and H. Yasuoka for discussions.
One of the authors (T.H.) is supported by Japan Society for
the Promotion of Science and by the Ministry of Education, Culture,
Sports, Science, and Technology of Japan.
The computation of this work has been done using the facilities
of Japan Atomic Energy Research Institute.


\end{document}